\documentstyle[sprocl,epsfig]{article}

\def\Journal#1#2#3#4{{#1} {\bf #2}, #3 (#4)}

\def\apj{{\em ApJ}}
\def\mnras{{\em MNRAS}}

\def\be{\begin{equation}}
\def\ee{\end{equation}}
\def\bea{\begin{eqnarray}}
\def\eea{\end{eqnarray}}

\def\rosat{{\it ROSAT}}
\def\asca{{\it ASCA}}
\def\ginga{{\it Ginga}}
\def\exosat{{\it EXOSAT}}
\def\tenma{{\it Tenma}}
\def\sl2{{\it Spacelab-2}}

\begin{document}

\title{REVISITING THE MASS OF THE COMA CLUSTER FROM X-RAY OBSERVATIONS}

\author{J.~P.~HUGHES}

\address{Department of Physics and Astronomy, Rutgers University,\\ 
P.~O.~Box 849, Piscataway, NJ 08855, USA\\
E-mail: jph@physics.rutgers.edu}


\maketitle\abstracts{I re-examine mass estimates of the Coma cluster
from pre-\asca\ X-ray spectral observations.  A large range of model
dark matter distributions are examined, under the assumptions of
hydrostatic equilibrium and spherical symmetry, to determine
the widest possible allowed range on the total mass of the 
cluster.  Within a radius of 1 Mpc, the total cluster mass is tightly 
constrained to be $(6.2 \pm 0.9) \times 10^{14}\, M_\odot$ and the
ratio of luminous baryonic matter to total matter lies between 13\%
and 17\%. Within a radius of 3 Mpc the total mass is $(1.3 \pm 0.5)
\times 10^{15}\, M_\odot$ and the luminous matter fraction is
20\%-40\%.  I find that the ``universal'' dark matter density profile
proposed by Navarro, Frenk, and White,\cite{nfw} based on N-body
simulations of a standard cold-dark-matter dominated Universe,
predicts a steep temperature gradient within the core of the cluster
that is a poor fit to the Coma data and can be rejected at greater
than 99\% confidence.}

\section{Introduction}

Determination of the total masses of galaxy clusters is of great
interest because of what these measurements can tell us
about the amount, nature, and distribution of dark matter.  The ratio
of luminous to dark matter, which provides a lower limit to the baryon
fraction, appears to be considerably higher in rich clusters of
galaxies than that allowed by Big-Bang nucleosynthesis in a flat
($\Omega=1$) $\Lambda=0$ Universe.  
The Coma Cluster, because it is nearby and well observed, stands as
an important laboratory for studies of this kind. 

I was motivated by several factors to revisit measuring the mass of
the Coma cluster from X-ray observations. One was to address the
reliability and accuracy of X-ray--derived cluster mass estimates,
which have been questioned recently.\cite{balland} Another was to see
if Coma's radial gas temperature profile, which is essential for
determining the gravitating mass, showed evidence for a remarkably
rapid drop at the cluster outskirts as seen in the cluster Abell 2163
by {\em ASCA}.\cite{markev} The availability of previously unpublished
scanning data from the \ginga\ satellite that provides strong
constraints on the temperature distribution over spatial scales up
to 1$^\circ$ was a further motivation.

\begin{figure}[t]
\begin{center}
\psfig{figure=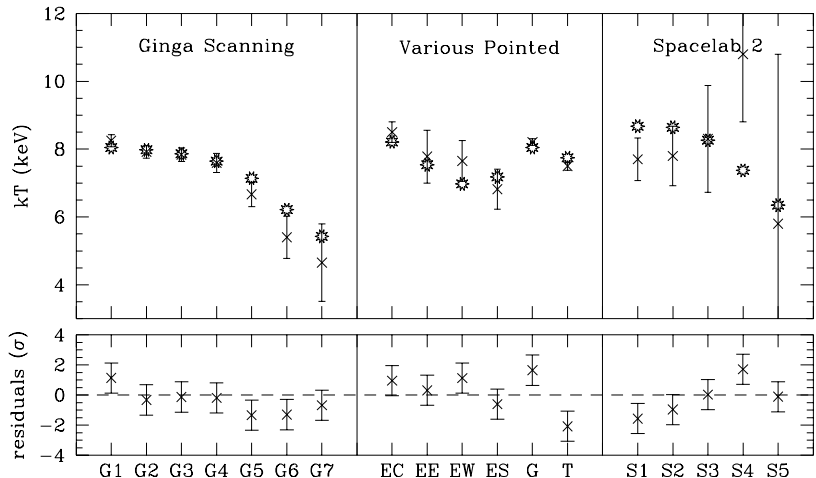,height=2.5in}
\end{center}
\caption{Projected temperatures of the hot gas in the Coma Cluster
(top panel) from various sources with one-sigma error bars (\ginga\
scan data on the left part of the plot; \sl2\ data on the right; and
pointed data from \tenma, \ginga, and several \exosat\ fields in the
middle). The temperature values from the overall best-fit dark matter
profile are shown as the starred symbols.  The bottom panel shows the
residuals between the data and the model.\label{fig:temp_prof}}
\end{figure}

In this article I utilize all available pre-\asca\ temperature data on
Coma from \tenma,\cite{tenma} \exosat,\,\cite{exosat}
\ginga\,\cite{ginga} and coded aperture mask data from
\sl2\,\cite{watt} (see Fig.~\ref{fig:temp_prof}). The \rosat\ surface
brightness profile~\cite{briel} was used to define the gas density
profile and the galaxy distribution was taken from Millington and
Peach.\cite{mill} I determine the cluster mass by constraining the
parameters of a dark matter halo function using methods described
previously.\cite{hughes} Briefly, given the \rosat\ gas density
profile and a particular set of parameter values for the dark matter
halo, one directly solves for the temperature profile, which is then
compared to the observed temperature values using a figure-of-merit
function (I employ $\chi^2$ here) to assess the quality of the fit. A
value of $H_0 = 50$ km s$^{-1}$ Mpc$^{-1}$, which yields a distance 
to Coma of 140 Mpc, is used throughout.

\section{Parameterized Dark Matter Halos}

\begin{figure}[t]
\begin{center}
\psfig{figure=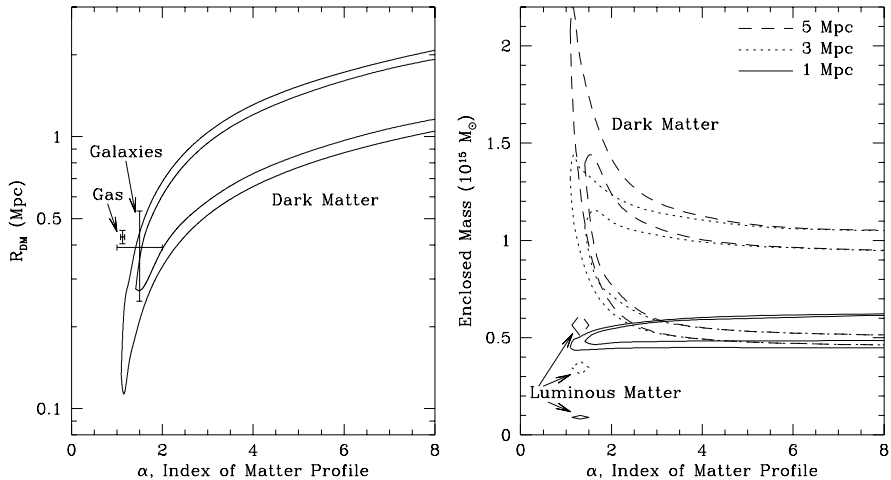,height=2.5in}
\end{center}
\caption{(Left) Allowed ranges (at the 90\% and 99\% confidence
levels) of the scale length of the dark matter distribution ($R_{\rm
DM}$) as a function of the power-law index of the matter distribution,
$\alpha$.  (Right) Allowed ranges (also 90\% and 99\% confidence
levels) of the masses of the luminous and dark matter in the Coma
cluster within three fiducial radii as indicated plotted as a function
of $\alpha$.\label{fig:encl_mass}}
\end{figure}

For the dark matter profile I use the three-parameter functional form
given by $\rho_{\rm DM} = \rho_0 [1+(R/R_{\rm DM})^2]^{-\alpha}$ and
the allowed ranges of all three parameters are determined. The fits to
the data are reasonably good (minimum $\chi^2 = 21.4$ for 15 degrees
of freedom; this model is plotted in Fig.~\ref{fig:temp_prof}) and are
summarized in Fig.~\ref{fig:encl_mass}. The left 
panel shows that the allowed values of the scale length $R_{\rm DM}$
and index $\alpha$ are strongly correlated. Formally I can reject a
dark matter model that is distributed like the X-ray gas with high
significance ($>$99\% confidence), but a model with the dark matter
distributed like the galaxies is allowed, although it is not highly
favored. Note that the current data do not constrain the maximum
value of $\alpha$ and, in fact, $\chi^2$ continues to drop, albeit
very slowly, for even larger values of $\alpha$ than shown.

The total gravitating mass of Coma is well constrained by these fits
as shown in the right panel of Fig.~\ref{fig:encl_mass}.  Here the
masses, integrated within three fiducial radii (1 Mpc, 3 Mpc, and 5
Mpc), were determined based on the allowed ranges of $R_{\rm DM}$,
$\alpha$, and $\rho_0$. (Note how the dark matter masses reach
asymptotic values at large values of $\alpha$ so, even though $\alpha$
is not constrained at the upper end, my mass estimates are robust.)
The total cluster mass (luminous plus dark) as well as the baryon
fraction within 1 Mpc and 3 Mpc are quoted in the abstract for the
99\% confidence level.  The mass within 5 Mpc is $(1.9 \pm 0.9) \times
10^{15}\, M_\odot$ and the luminous matter fraction is 20\%--55\%. 

The temperature profiles we obtain are nearly all convectively stable,
i.e., $\vert {d \ln T \over d \ln \rho}\vert < {2\over 3}$, within the
maximum observed extent of the cluster $\sim$3.3 Mpc. The few models
that violate this constraint lie between the 90\% and 99\% confidence
contours at low values of $R_{\rm DM}$ and $\alpha$ and thus are
statistically less likely to be acceptable solutions anyway.  In
summary, I find no evidence for steep temperature gradients in Coma
that might indicate nonhydrostatic or other exotic conditions.

\section{``Universal'' Dark Matter Halos}

Based on their $N$-body simulations of a standard cold-dark-matter
dominated Universe, Navarro, Frenk, \& White~\cite{nfw} (NFW) find
that the radial dark matter density profiles of systems ranging from
dwarf galaxies to rich clusters of galaxies can be well described by
the simple function 
$\rho = \delta_c \rho_{\rm crit}/(r/r_s)(1+r/r_s)^2$, 
where $\delta_c$ is the characteristic overdensity of the halo in
terms of the critical density $\rho_{\rm crit}$ and $r_s$ is a
characteristic scale radius.

The best fit of this function to the Coma temperature data is obtained
for $r_s \sim 0.5$ Mpc and $\delta_c \rho_{\rm crit} \sim 4 \times
10^{-26}$ g cm$^{-3}$.  However, this fit is formally unacceptable:
the minimum $\chi^2$ of 31.5 for 15 degrees of freedom can be rejected
at greater than 99\% confidence. The central temperature in the
best-fit model is rather high $\sim$16.5 keV but drops rapidly to
$\sim$8 keV at $r_s$ and thereafter continues to fall, although more
gradually.  The steep temperature gradient near the center of the
cluster is inconsistent with the observed data for Coma.  This also
appears to be the case for the rich cluster Abell 2256 where the
NFW halo function predicts a considerably steeper temperature
profile in the center of the cluster than observed.\cite{kop}  On the
other hand, the NFW function appears to be an acceptable description
of the optically-derived average mass profile of galaxy
clusters.\cite{carl} Further work on the mass distribution of Coma as
well as other galaxy clusters is clearly needed in order to resolve
this important issue.


\par\noindent
{\small This research was partially supported by NASA LTSA Grant NAG5-3432.}

\end{document}